\documentclass[%
 reprint,
superscriptaddress,
 amsmath,amssymb,
 aps,
 pre,
]{revtex4-1}

\usepackage{graphicx}
\usepackage{dcolumn}
\usepackage{bm}
\usepackage{multirow}

\begin{document}


\title{Extended self-similarity of atmospheric boundary layer wind fields\\ in mesoscale regime: Is it real?}

\author{V. P. Kiliyanpilakkil}
\author{S. Basu}%
\email{sukanta\_basu@ncsu.edu}
\affiliation{%
 Department of Marine, Earth, and Atmospheric Sciences, North Carolina State University, Raleigh, NC 27695, USA}

\date{\today}

\begin{abstract}
In this letter, we study the scaling properties of multi-year observed and atmospheric model-generated wind time series. We have found that the extended self-similarity holds for the observed series, and remarkably, the scaling exponents corresponding to the meoscale range closely match the well-accepted inertial-range turbulence values. However, the scaling results from the simulated time series are significantly different.
\end{abstract}

\maketitle

In   the  turbulence  literature,   the  scaling   exponent  spectrum,
$\zeta_p$, is defined as:
\begin{equation}
S_p(r) = \langle |\Delta u|^p \rangle \sim r^{\zeta_p}
\end{equation}
where $S_p(r)$  is the $p$-th order structure  function. The  angular bracket denotes spatial averaging  and $r$  is a separation distance that varies  within a specific scaling range (e.g., inertial-range). For time series analysis (where Taylor's hypothesis is inapplicable), the usage of time increment, $\Delta t$ (in lieu of $r$), and temporal averaging is customary. According to Kolmogorov's celebrated 1941 hypothesis (K-41; \cite{kolmogorov1941local}), $\zeta_p$ equals to $p/3$ in the isotropic inertial-range of turbulence. 

More than two decades ago, Benzi and co-workers \cite{benzi1993extended} proposed the 
extended self-similarity (ESS) framework for the characterization of 3D fully developed turbulence. In this framework, structure functions of different orders are plotted against one another for the identification of scaling regimes, and, more importantly, for the robust estimation of the relative scaling exponents ($\zeta_{p,q}^* = \zeta_p/\zeta_q$). Over the years, numerous studies \cite{benzi1993EPL,benzi1994scaling,benzi1995PhysD,benzi1995intermittent,arneodo1996structure} have demonstrated the strength of ESS in terms of identifying scaling regimes even when the traditional structure function approach fails. Theoretical justification for the existence of ESS was provided by Benzi et al. \cite{benzi1996scaling}, and more recently, by Chakraborty et al. \cite{chakraborty2010extended}.

Thus far, most of the ESS-related studies, in the arena of turbulence, focused on the longitudinal component of velocity fields. However, a handful of studies reported ESS to be applicable to the transverse component of velocity \cite{antonia1997scaling}, as well as for temperature fields \cite{antonia1997scaling,skrbek2002temperature}. Furthermore, numerically generated turbulence data were also found to follow ESS reasonably well (e.g., \cite{benzi1994scaling,grossmann1997application}). 
It is important to note that, in the presence of strong shear, the applicability of ESS seems to be somewhat limited \cite{amati1997extended,ruiz2000scaling}. The so-called generalized ESS (or G-ESS) framework \cite{benzi1996scaling,biferale1997generalized} is more appropriate for this scenario. 

Beyond the turbulence research community, the ESS framework has been embraced by many researchers from diverse disciplines: geology \cite{nikora2001extended}, biology \cite{atyabi2006two}, image processing \cite{turiel1998self}, finance \cite{constantin2005volatility},  to name a few. In this respect, the concluding remark by  Nikora and Goring \cite{nikora2001extended} is noteworthy: 
\begin{quote}
``Finally, our considerations suggest that, indeed ESS and G-ESS may be an inherent property of many natural phenomena rather than a property exclusively of turbulence.'' 
\end{quote}

In a recent work, Kiliyanpilakkil et al.~\cite{K15} analyzed long-term wind speed time series from several field sites around the world with diverse meteorological and geographical conditions. They reported that the wind fields in the lower part of the atmospheric boundary layer (specifically, up to a height of 300 m from the ground) remarkably follow ESS. The scaling regime was found to be within the range of ten minutes to six hours (called the mesoscale range), far beyond the inertial-range of turbulence. Most intriguingly, the relative scaling exponents were estimated to be marginally different (more intermittent) from the commonly reported inertial-range values (e.g.,  \cite{frisch1995turbulence}).

\begin{figure*}[ht]
\begin{center}
\includegraphics[width=2.3in]{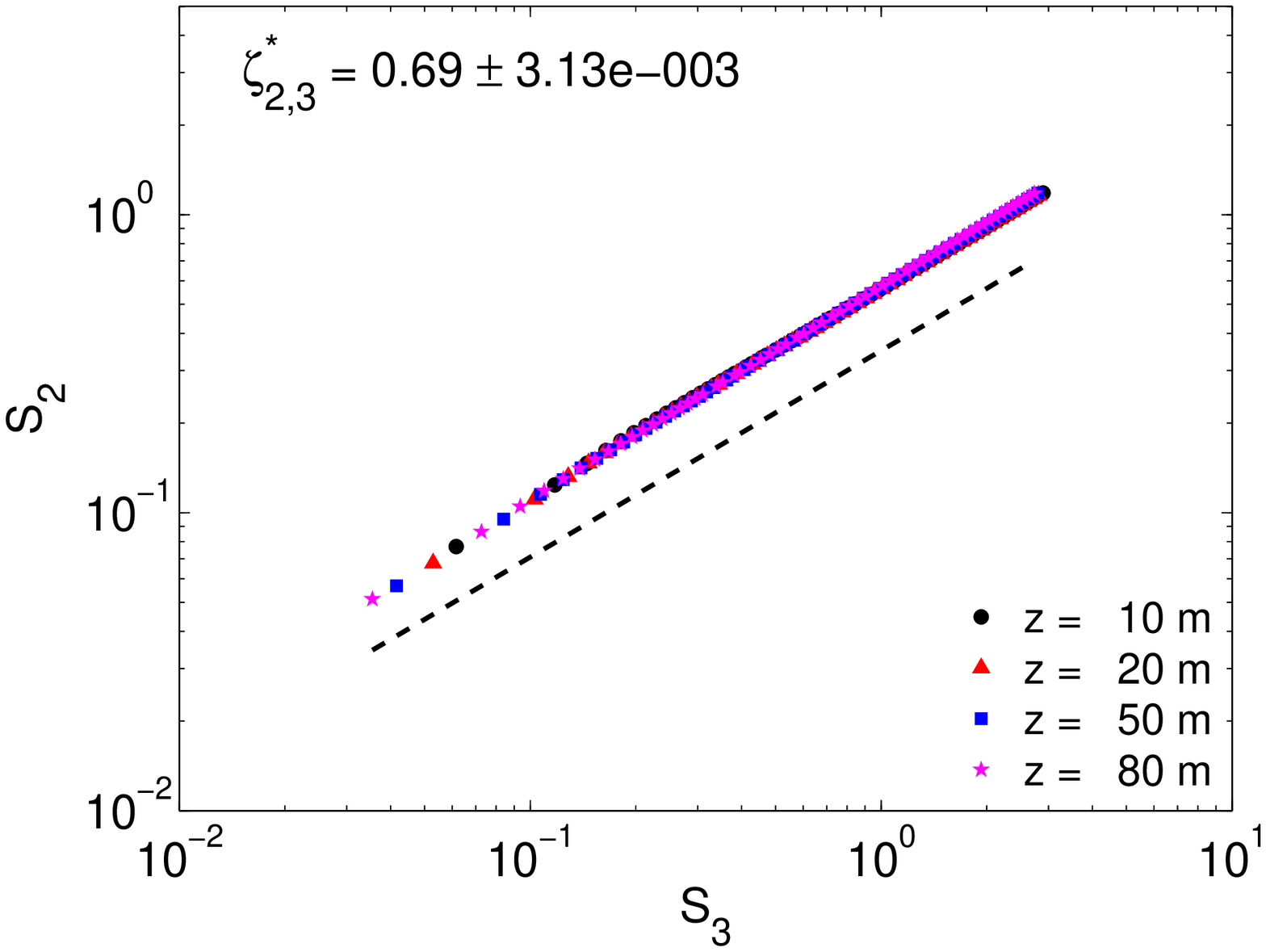}
\includegraphics[width=2.3in]{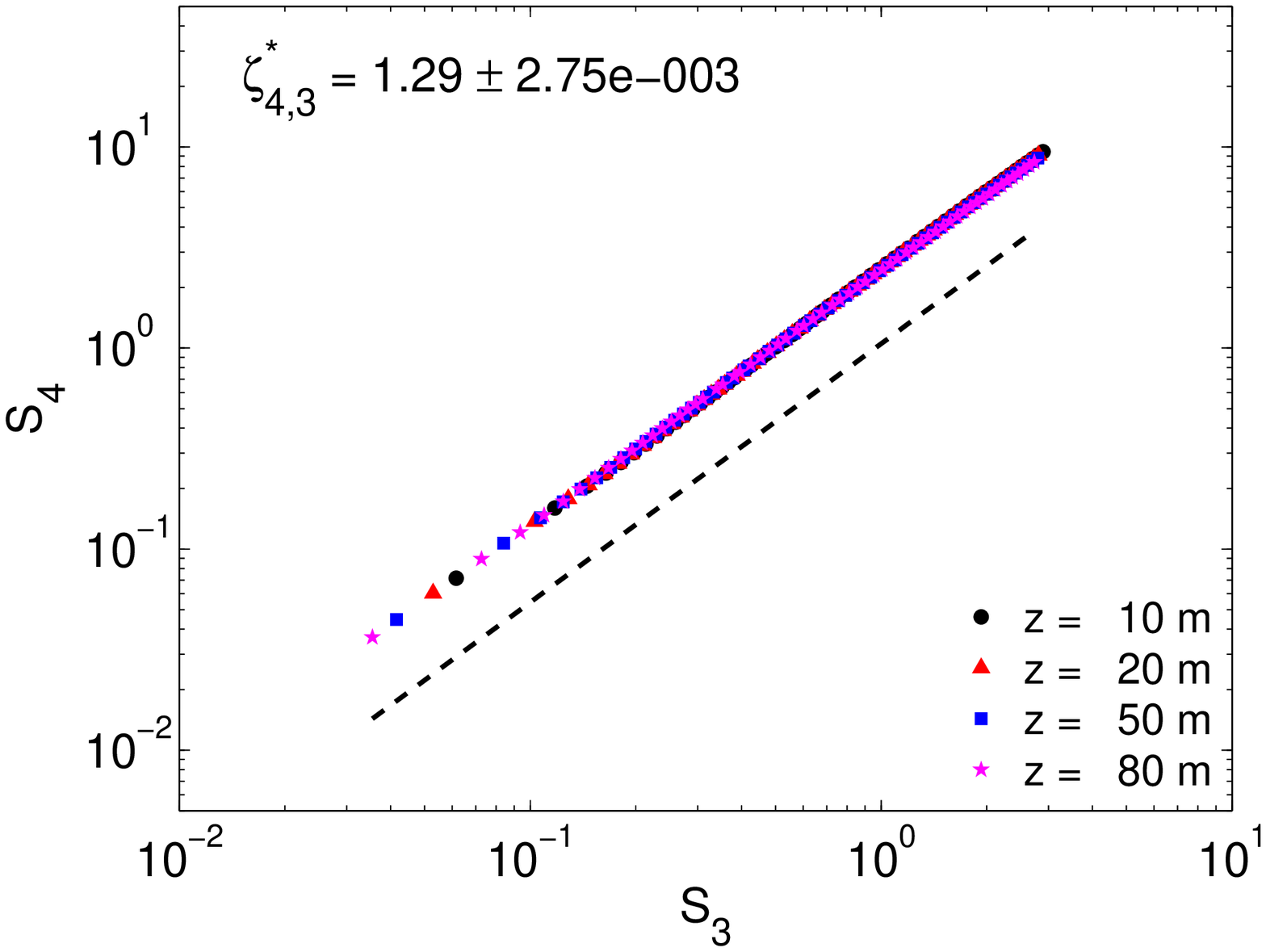}
\includegraphics[width=2.3in]{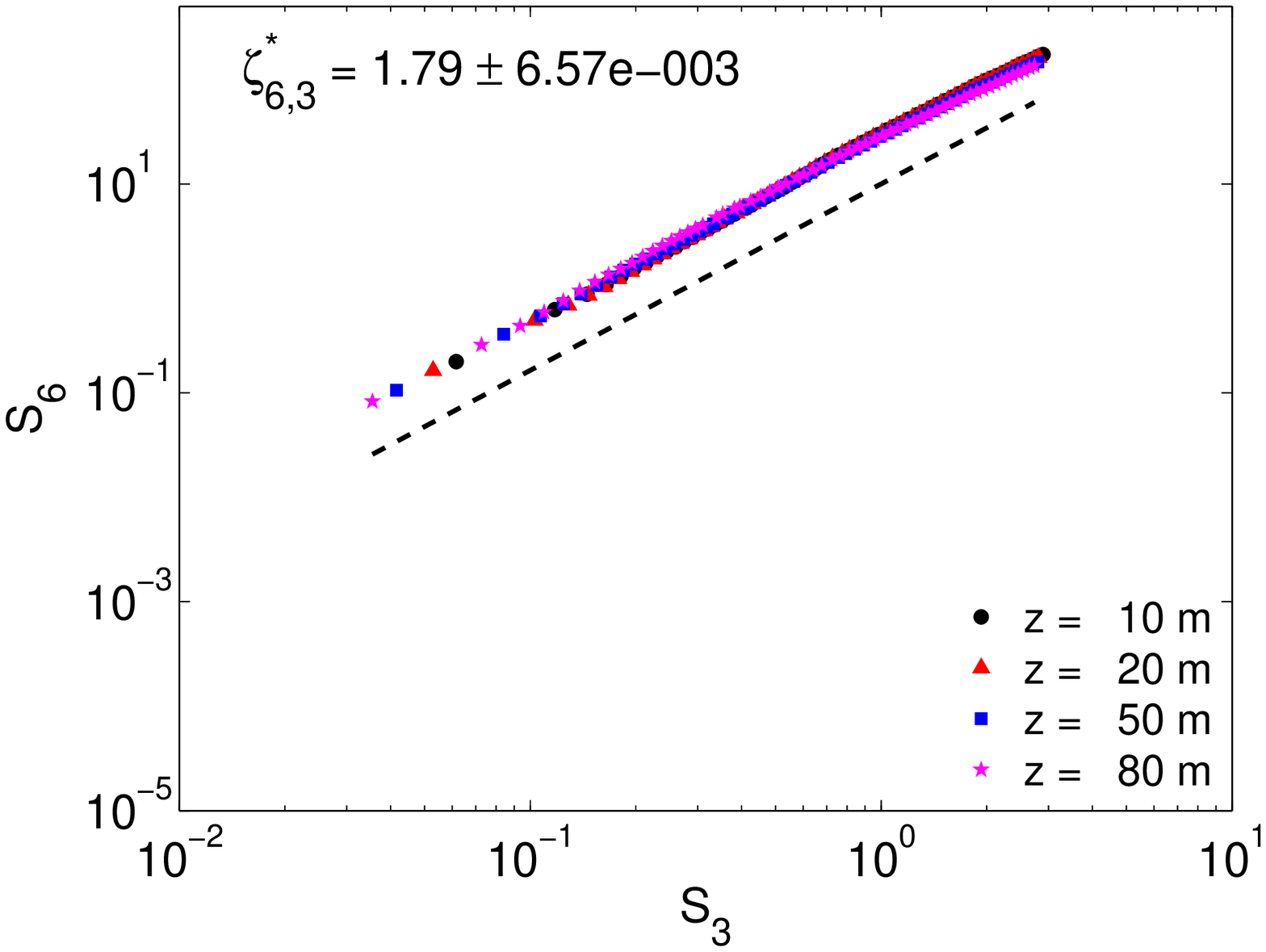}\\
\includegraphics[width=2.3in]{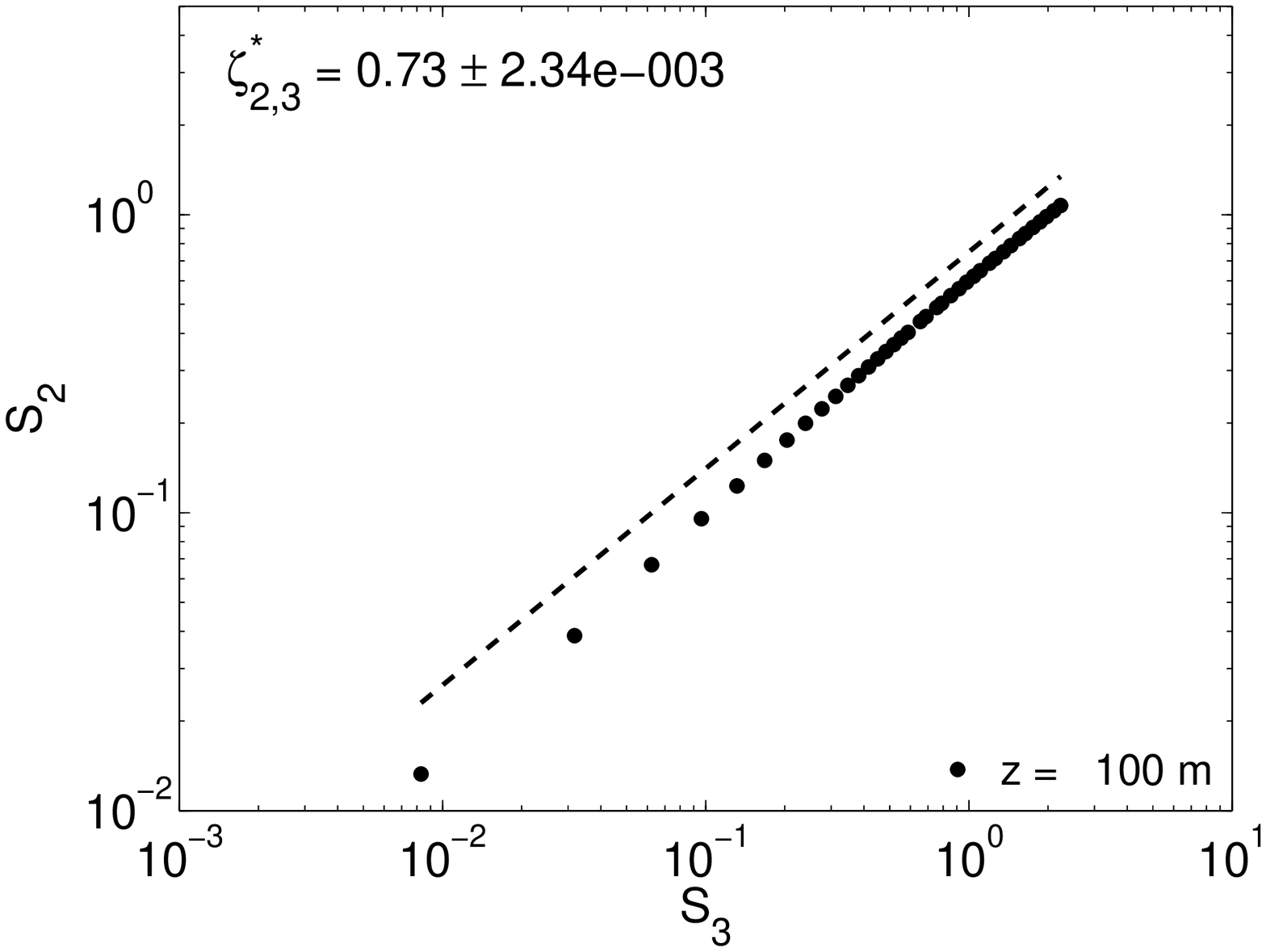}
\includegraphics[width=2.3in]{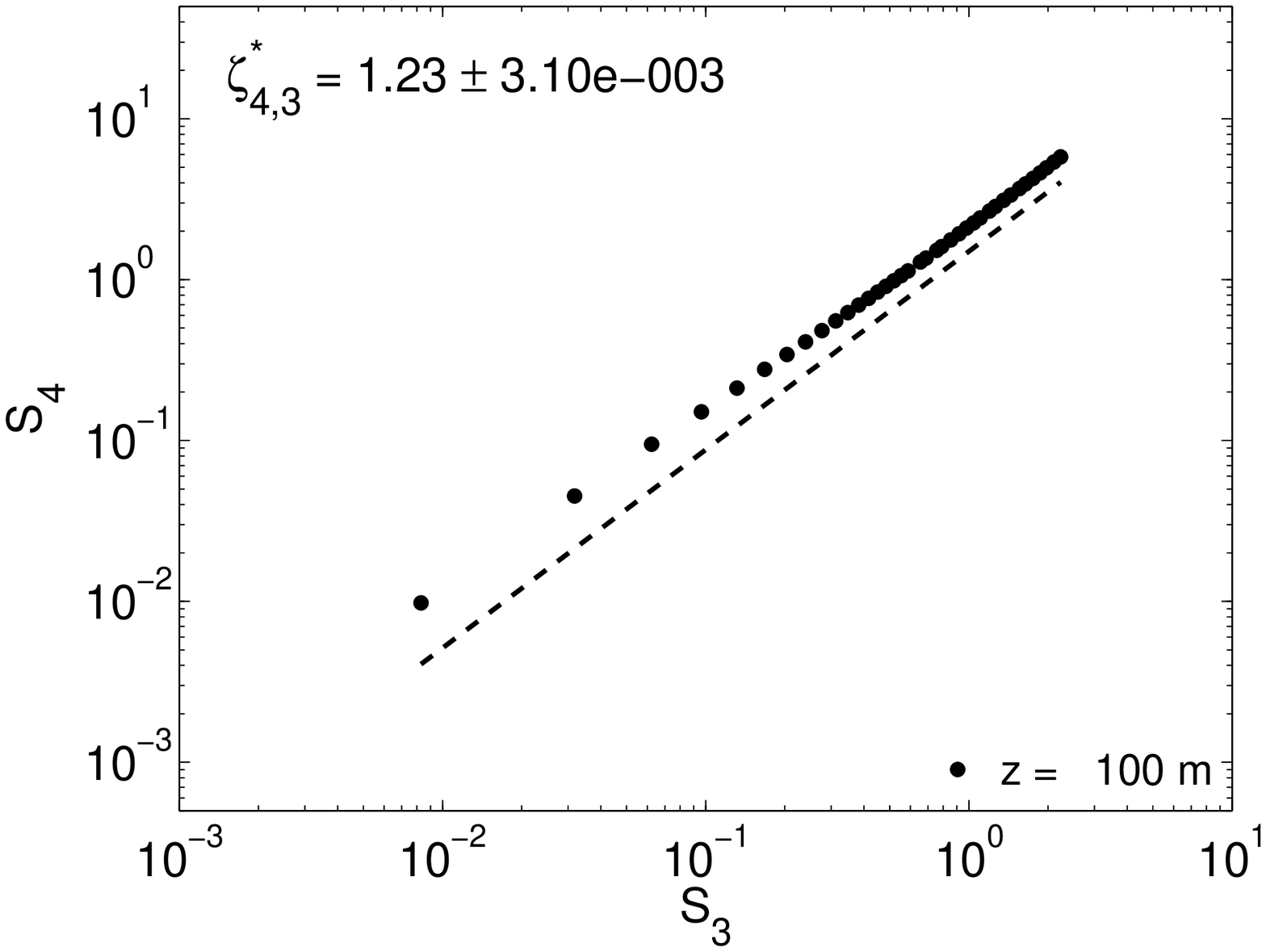}
\includegraphics[width=2.3in]{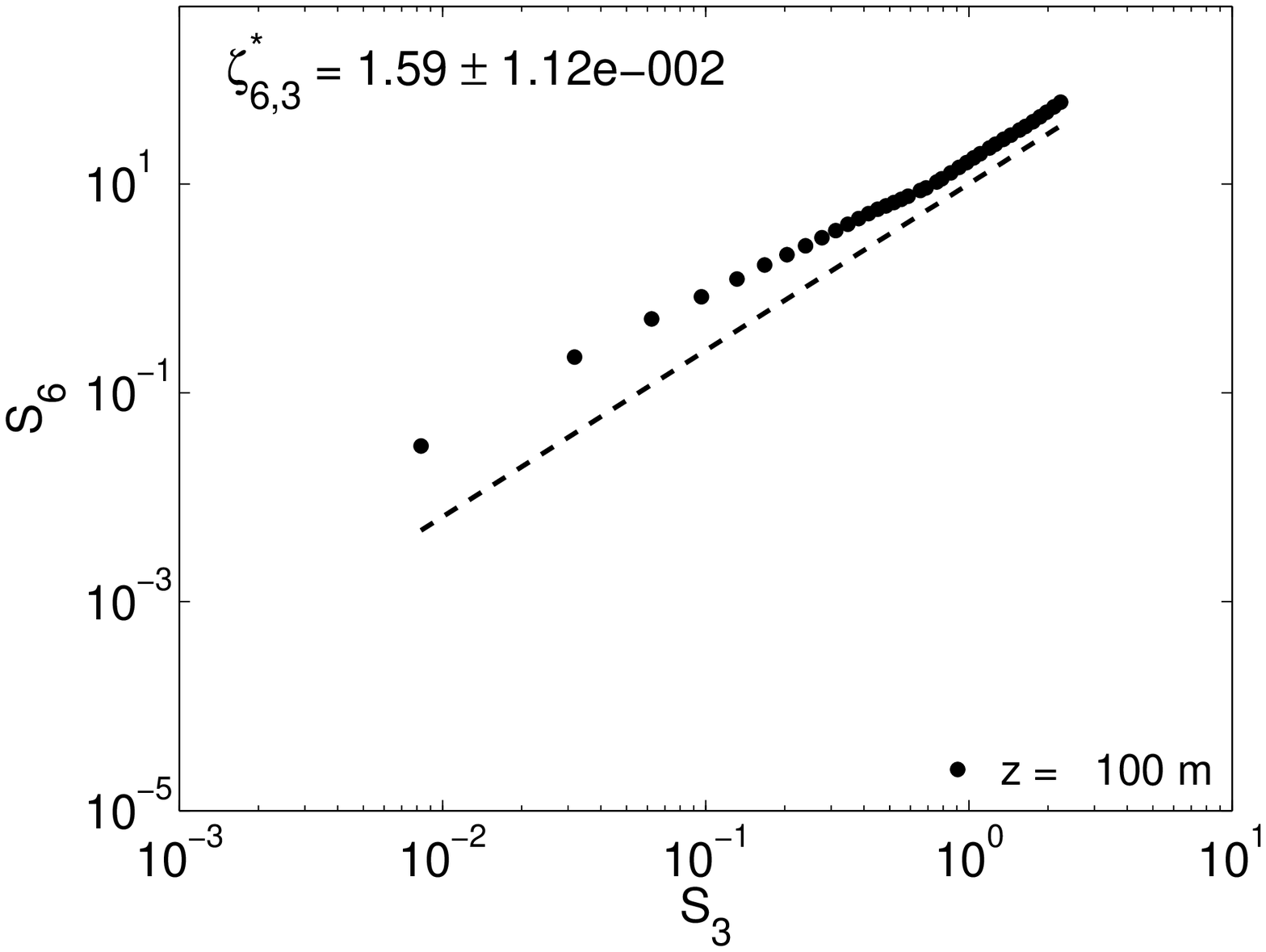}\caption{The variation of the second-order (left panel), fourth-order (middle panel) and sixth-order (right panel) structure functions with respect to the third-order structure functions. Observed (top row) and simulated (bottom row) wind speed time series are used to create these plots. The relative scaling exponents ($\zeta^*_{p,q}$) are reported on the top-left corner of each plots. The mean and standard deviation of $\zeta^*_{p,q}$ are estimated via bootstrapping. The dashed line in each plot represents the mean value of $\zeta^*_{p,q}$. In the top row plots, structure function values corresponding to $\Delta t = $ 1 min -- 6 h are used for $\zeta^*_{p,q}$ estimation. Whereas, in the bottom row plots, $\zeta^*_{p,q}$ values correspond to only $\Delta t = $ 2--6 h.}
\label{f1}
\end{center}
\end{figure*}

In the present letter, we examine whether the wind fields near complex terrain also follow ESS in the mesoscale regime. Since Kiliyanpilakkil et al.~\cite{K15} only considered field sites with homogeneous, flat surface (including an offshore site), it is imperative to document if such idealized conditions are indeed pre-requisites for the existence of ESS. Next, we investigate whether a state-of-the-art atmospheric model with (imperfect) physical parameterizations can capture the ESS-based scaling traits of observed wind fields. If it does, then, the ESS results could be considered trivial from an atmospheric modeling perspective. Otherwise, these non-trivial scaling results might be utilized as benchmarks for atmospheric models. 

Wind speed and other meteorological data from the 80 m tall M2 tower (http://www.nrel.gov/midc/nwtc\_m2/), near Boulder, Colorado (latitude: 39.9107 N; longitude: 105.2348 W) have long been utilized by the wind energy research community. This tower is maintained by the National Wind Technology Center (NWTC) and the wind sensors are calibrated annually \cite{johnson2000design}. In this study, we make use of long-term (years 2004--2014) wind speed time series measured at the heights of 10 m, 20 m, 50 m, and 80 m using cup anemometers. Relatively short averaging time (1 min), large sample size ($\approx 5.78$ millions for each time series), and virtually no data gaps (only 89 missing samples in each time series) make these time series highly desirable for scaling analyses. Furthermore, the location of the M2 tower is ideal for the present study. Being in the lee of the Colorado Rocky mountains, this location is prone to quite complex wind flows \cite{kell06}. 

\begin{table*}[ht]
\caption{ESS-based relative scaling exponents (mean $\pm$ standard deviation) estimated via bootstrapping.}
\label{T1}
\begin{center}
\begin{tabular}{clccc}
\hline
Height (m)  & Variables & $\zeta^*_{2,3}$ & $\zeta^*_{4,3}$ & $\zeta^*_{6,3}$\\
\hline
\multirow{3}{*}{10} & Wind &	0.69	$\pm$	3.13e-003	&	1.29	$\pm$	2.71e-003	&	1.79	$\pm$	6.56e-003\\
 & Surrogate (FT) &	0.67	$\pm$	2.53e-005	&	1.33	$\pm$	4.29e-005	&	2.00	$\pm$	1.67e-004\\
 & Surrogate (IAAFT) &	0.68	$\pm$	8.51e-004	&	1.31	$\pm$	1.44e-003	&	1.91	$\pm$	5.33e-003\\
\hline
\multirow{3}{*}{20} & Wind &	0.70	$\pm$	3.10e-003	&	1.28	$\pm$	2.84e-003	&	1.78	$\pm$	8.31e-003\\
 & Surrogate (FT) &	0.67	$\pm$	2.43e-005	&	1.33	$\pm$	4.98e-005	&	2.00	$\pm$	2.48e-004\\
 & Surrogate (IAAFT) &	0.68	$\pm$	8.62e-004	&	1.31	$\pm$	1.54e-003	&	1.91	$\pm$	5.84e-003\\
\hline
\multirow{3}{*}{50} & Wind &	0.71	$\pm$	2.20e-003	&	1.26	$\pm$	1.47e-003	&	1.73	$\pm$	6.52e-003\\
 & Surrogate (FT) &	0.67	$\pm$	2.19e-005	&	1.33	$\pm$	4.45e-005	&	2.00	$\pm$	2.07e-004\\
 & Surrogate (IAAFT) &	0.68	$\pm$	8.46e-004	&	1.31	$\pm$	1.53e-003	&	1.91	$\pm$	5.97e-003\\
\hline
\multirow{3}{*}{80} & Wind &	0.72	$\pm$	1.40e-003	&	1.26	$\pm$	1.29e-003	&	1.68	$\pm$	1.23e-002\\
 & Surrogate (FT) &	0.67	$\pm$	1.10e-005	&	1.33	$\pm$	2.24e-005	&	2.00	$\pm$	1.02e-004\\
 & Surrogate (IAAFT) &	0.68	$\pm$	8.20e-004	&	1.31	$\pm$	1.44e-003	&	1.92	$\pm$	5.60e-003\\
\hline
\end{tabular}
\end{center}
\end{table*}

Prior to scaling analysis, we normalize (zero mean, unit variance) each time series. In the top row of Fig.~\ref{f1}, using the ESS framework, we report various structure functions corresponding to $\Delta t = $ 1 min -- 6 h. Undoubtedly, the ESS holds for all the wind time series from different altitudes. The collapse of the data points on a single curve is truly remarkable. The relative exponents ($\zeta^*_{p,q}$) are estimated using bootstrapping, a popular resampling technique \cite{efron1982jackknife,mooney1993bootstrapping,good06}. In this approach, $N$ number of resamples are randomly drawn with replacement from an original sample of size $N$. This drawing operation is repeated numerous times (a Monte-Carlo procedure) -- in this study, we use 10,000 random drawings for each case. From each drawing, one value of $\zeta^*_{p,q}$ is calculated using ordinary least-squares approach. The mean and standard deviation of $\zeta^*_{p,q}$ for each sensor height are reported in Table~\ref{T1}. In addition, the relative scaling exponents for the aggregated set is reported on the top-left corner of each plot. It is needless to point out that the estimated $\zeta^*_{p,q}$ are very close to the corresponding inertial-range values \cite{frisch1995turbulence}. We detect a minute systematic dependence of $\zeta^*_{p,q}$ with height (see Table~\ref{T1}). Given that the overall variation of $\zeta^*_{p,q}$ is much less than 10\%, we deemed it unnecessary to invoke G-ESS.

In the bottom row of Fig.~\ref{f1}, we report the ESS results for a simulated wind series (years 2007--2012) from a grid-point close to the M2 tower. This simulated series was extracted from an unparalleled wind database, called the Wind Integration National Dataset (WIND) Toolkit, recently created by the National Renewable Energy Laboratory \cite{draxl2015wind}. A state-of-the-art atmospheric model, known as the Weather Research and Forecasting (WRF) model \cite{Skamarocketal2008} was the workhorse behind these computationally challenging simulations. In this model, most of the atmospheric processes (e.g., turbulence, radiation, microphysics, land-atmosphere interactions) are parameterized. The innermost computational domain covered the entire United States with horizontal grid spacing of 2 km. Please refer to Draxl et al.~\cite{draxl2015wind} for other numerical configurations and physical parameterization settings. In the WIND Toolkit, wind speed data at the 100 m above ground level are available every 5 min. Thus, for a selected grid point, for the period 2007--2012, the total sample size is $\approx$ 0.6 millions. 

The ESS analyses of the simulated wind series are shown in the bottom row of Fig.~\ref{f1}. For longer time-scales ($\Delta t > $ 2 h), the simulated results are in agreement with ESS. The estimated $\zeta^*_{p,q}$ for $\Delta t = $ 2--6 h are close (slightly more intermittent) to the observed values reported in Table~\ref{T1}. However, for $\Delta t = $ 5 min -- 2 h, the simulated results strongly deviate from ESS. From atmospheric boundary layer wind modeling standpoint, the turbulence closure scheme is the most relevant physical parameterization. Thus, we speculate that the usage of a first-order closure scheme (called the Yonsei University -- YSU scheme \cite{hong2010new}) in the creation of the WIND Toolkit is at the root of this anomaly. In the near-future, we will be exploring if any higher-order closure scheme has the ability to capture the ESS-based scaling characteristics in a more faithful manner. 

In the presence of noise and/or due to limited sample size, the estimation of accurate scaling exponents from time series is a challenging task. Thus, in order to assess the statistical significance of the aforementioned results, we employ the surrogate data-based hypothesis testing approach \cite{theiler1992testing,schreiber2000surrogate}. We synthetically generate two types of surrogates. The Fourier Transform phase randomized surrogates (henceforth FT) preserve the spectra (or, linear correlation structure) of a given time series \cite{theiler1992testing}. The other type of surrogate is created by the Iterative Amplitude Adjusted Fourier Transform (IAAFT) technique \cite{schreiber1996improved}. The IAAFT surrogates not only match the spectra of the original series (almost) perfectly, they also completely preserve its probability density function. 
One hundred realizations for each type of surrogate are generated for the simulated wind series. Due to high computational cost associated with the IAAFT algorithm, the number of realizations are reduced to ten for the observed series (each with a large sample size of $\approx$ 5.8 millions). To avoid any periodicity artifacts, during the synthesis of surrogates, we utilized the end point mismatch reduction strategy recommended by Schreiber and Schmitz \cite{schreiber2000surrogate}.  

\begin{figure*}
\begin{center}
\includegraphics[width=2.3in]{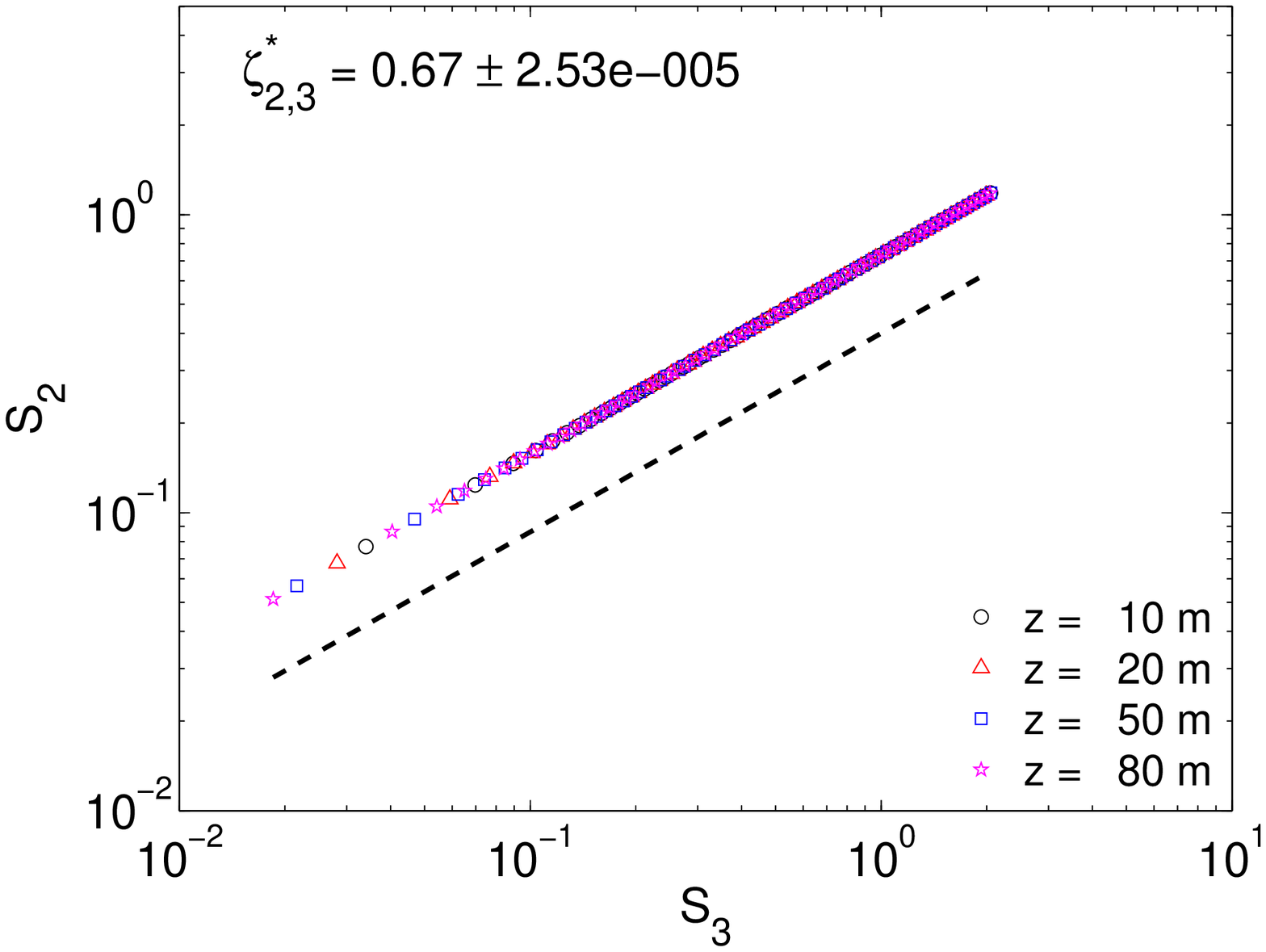}
\includegraphics[width=2.3in]{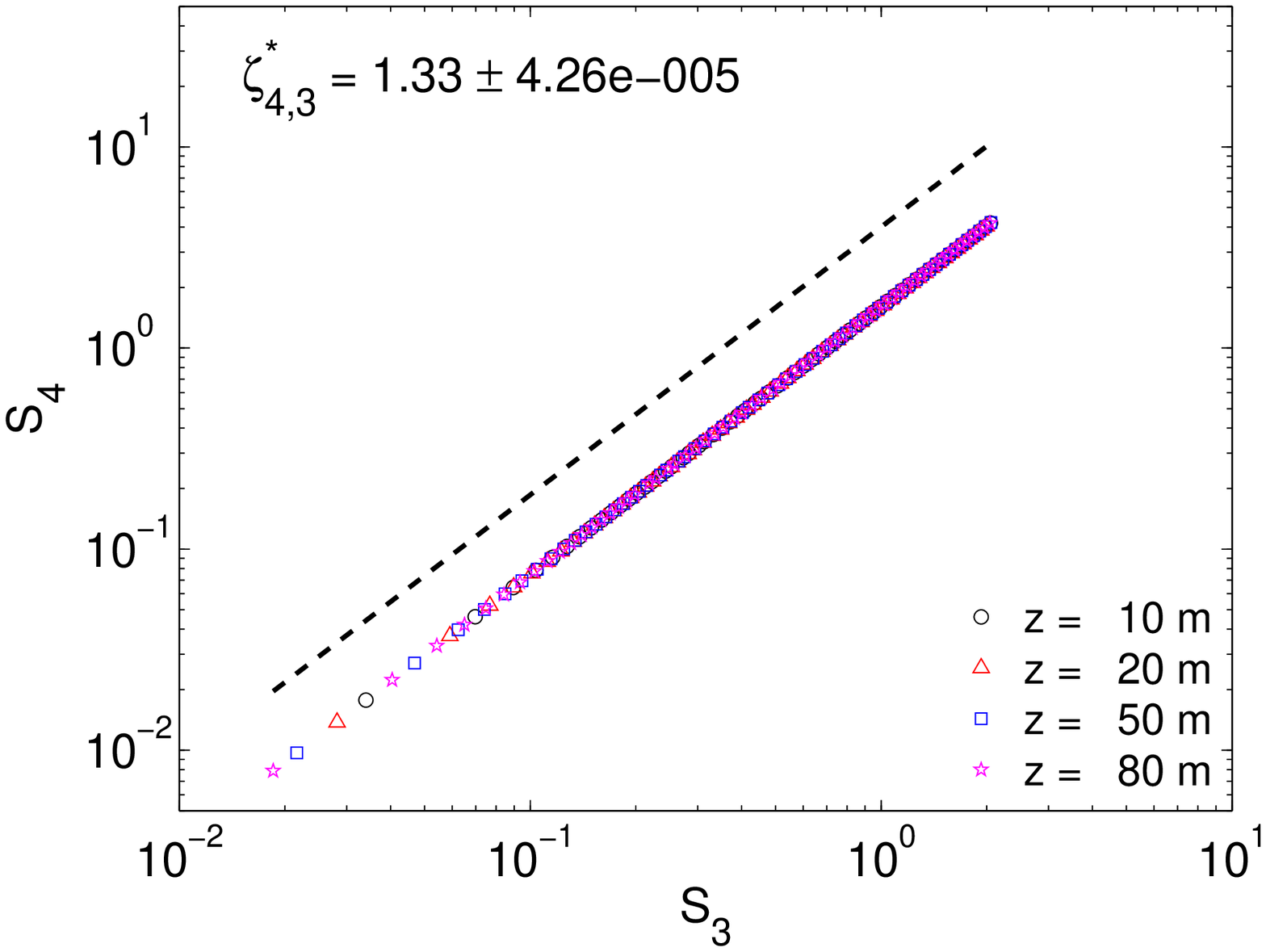}
\includegraphics[width=2.3in]{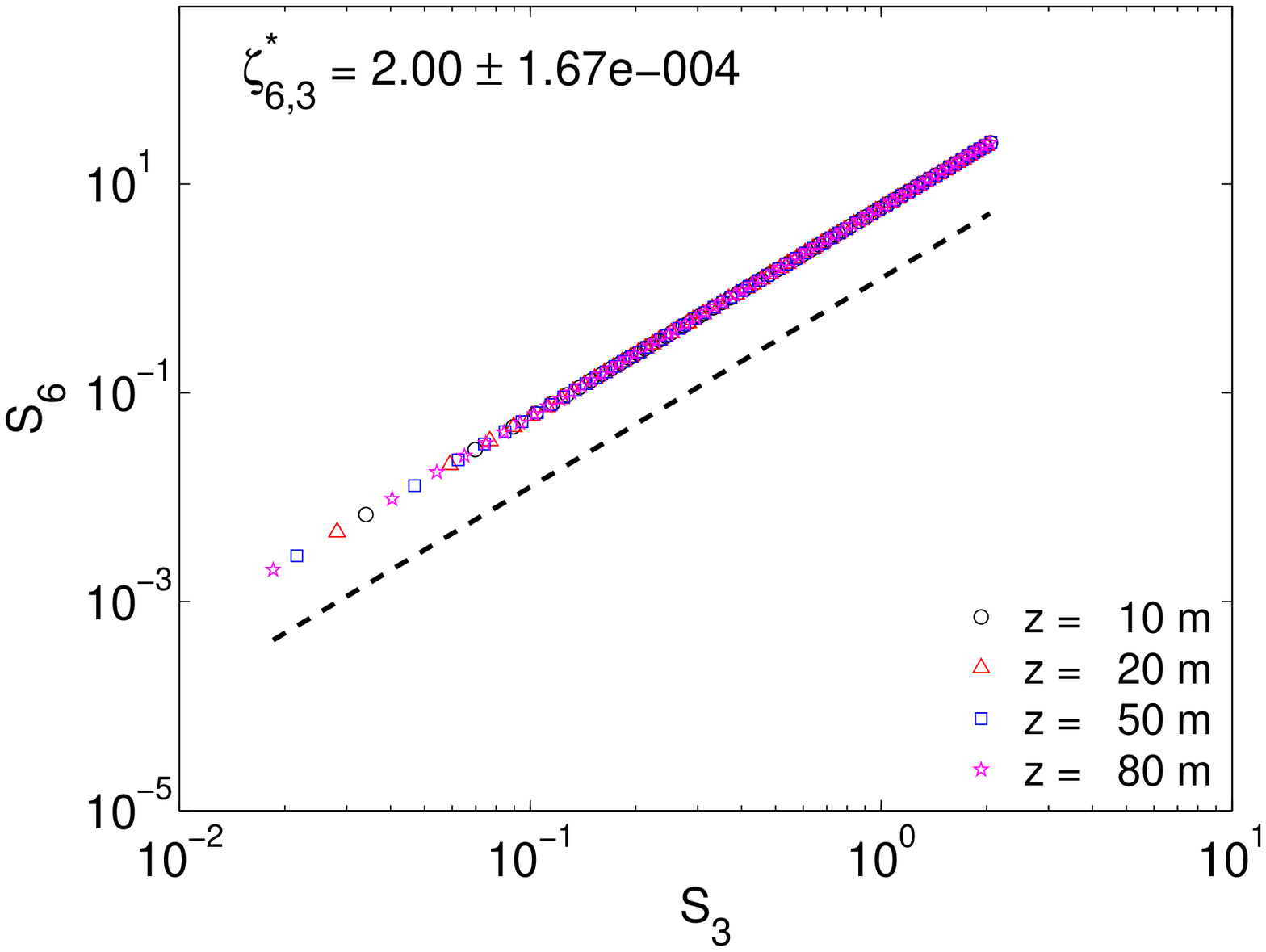}\\
\includegraphics[width=2.3in]{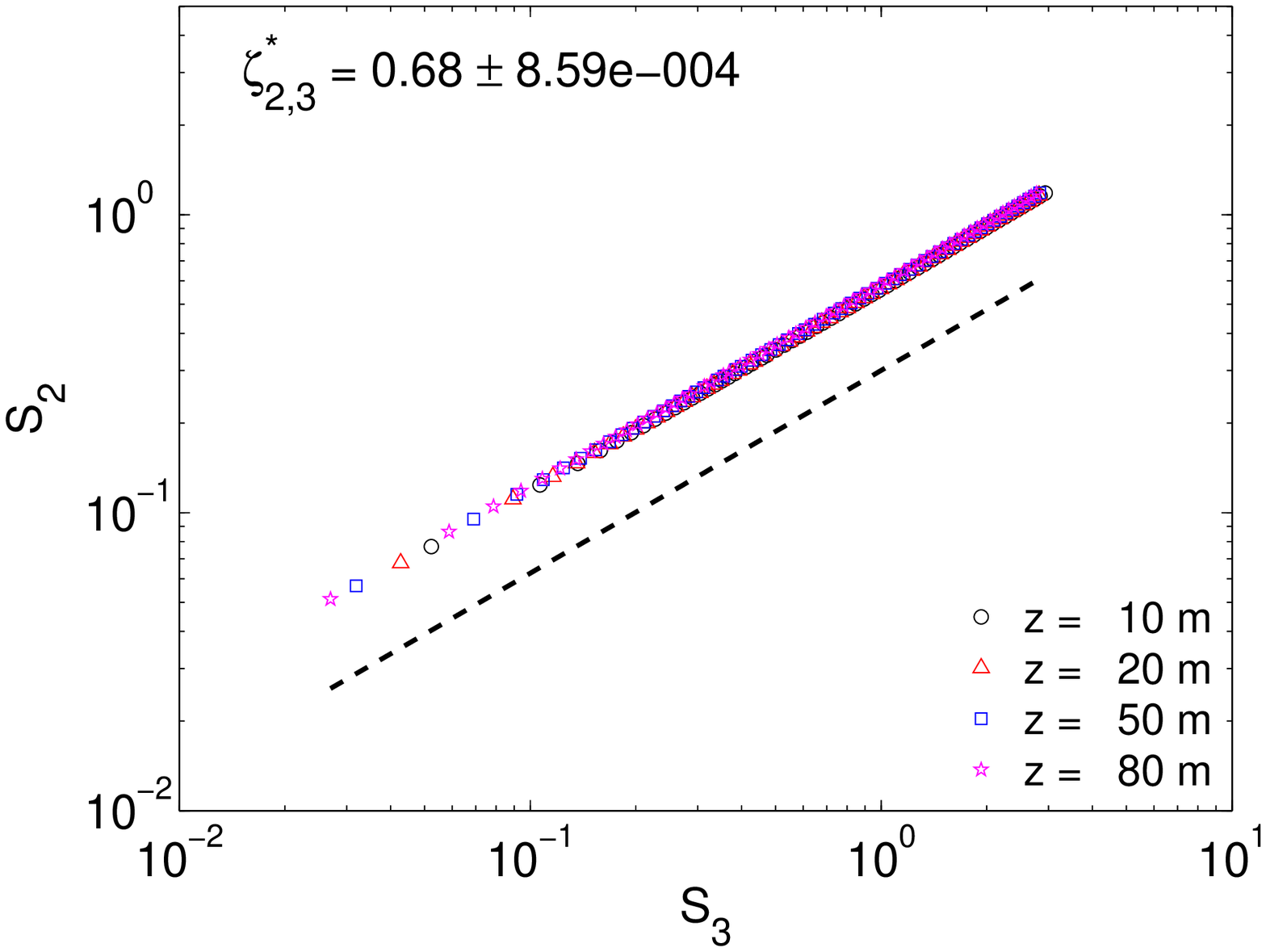}
\includegraphics[width=2.3in]{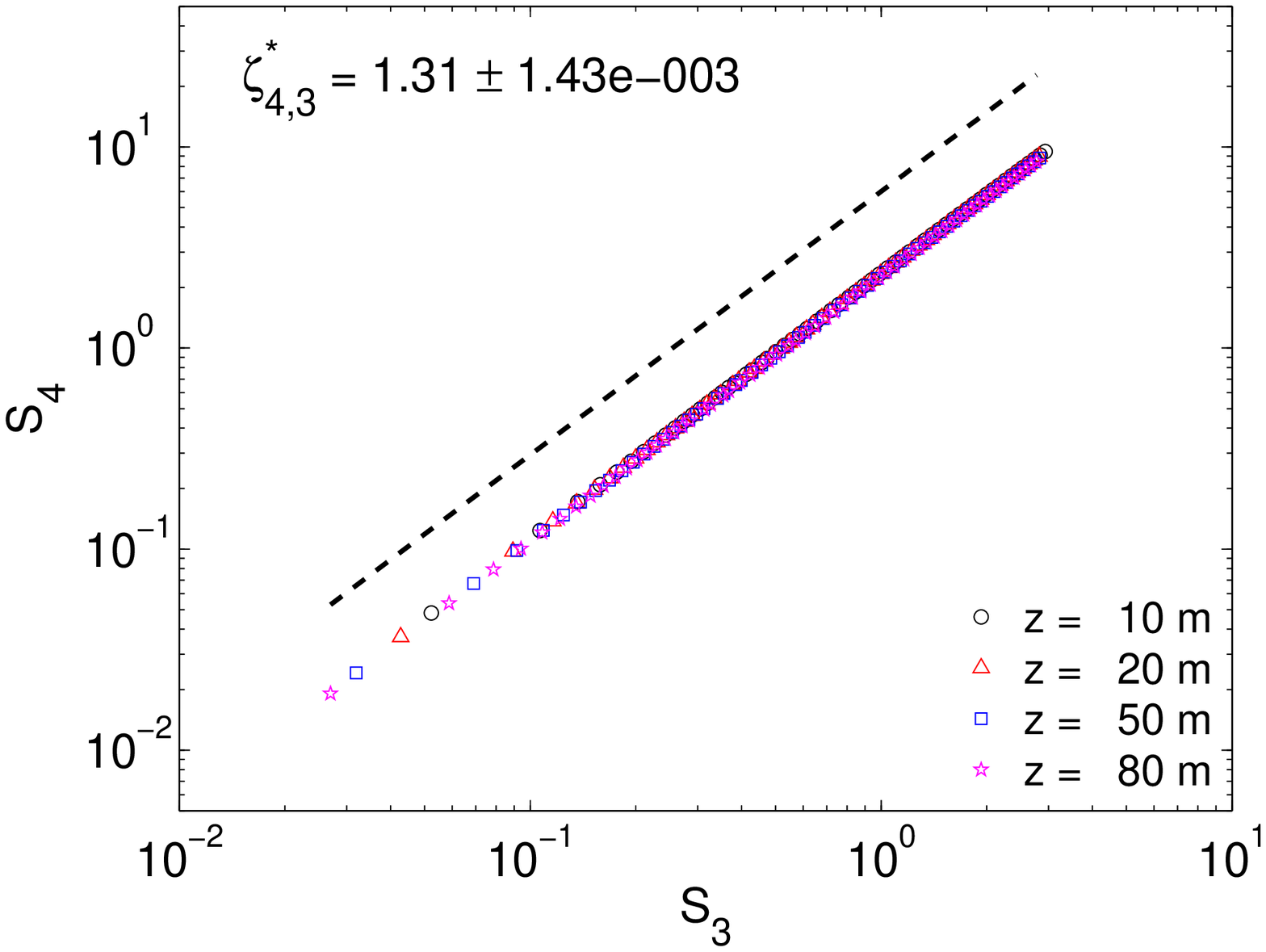}
\includegraphics[width=2.3in]{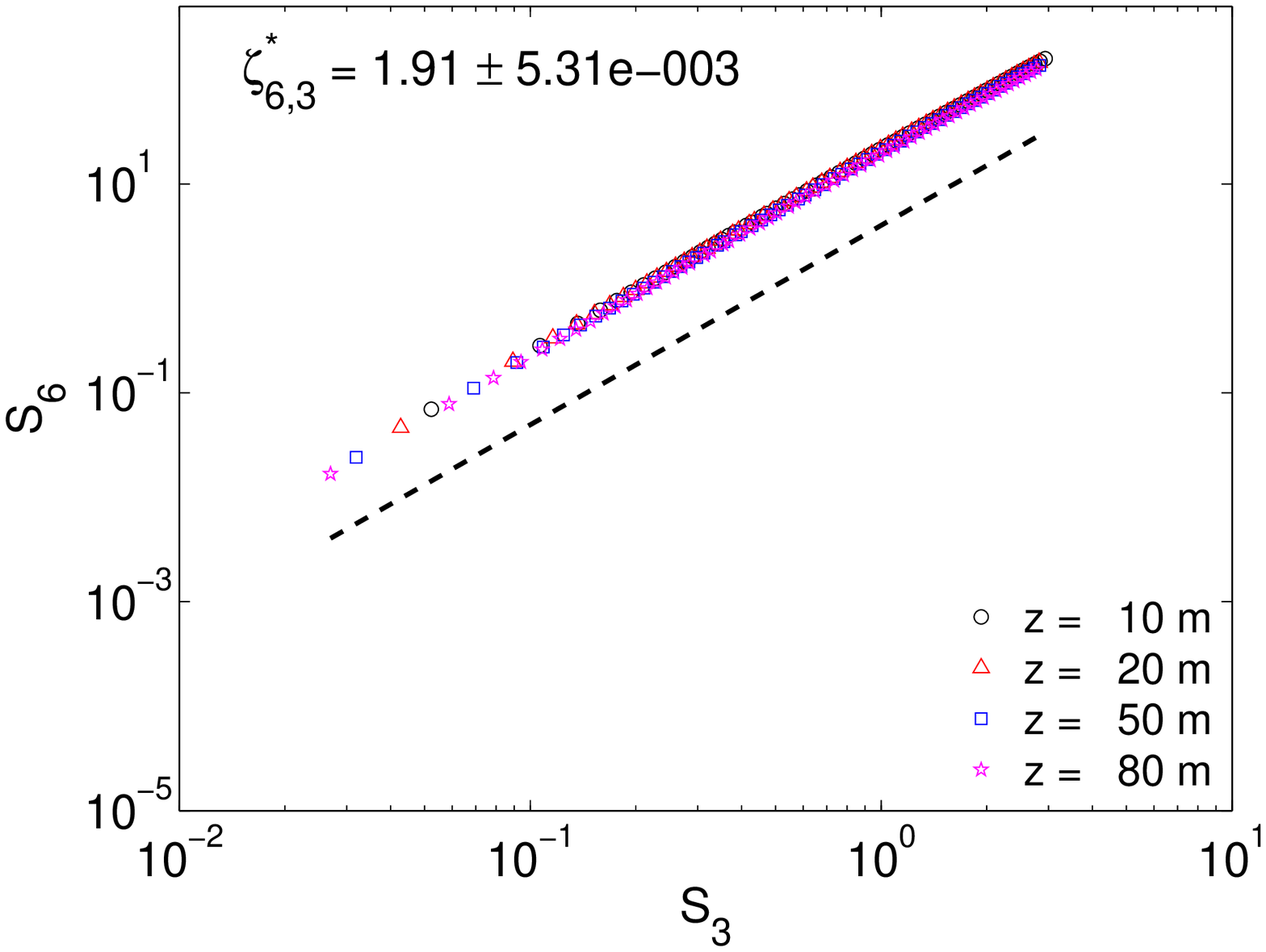}\\
\includegraphics[width=2.3in]{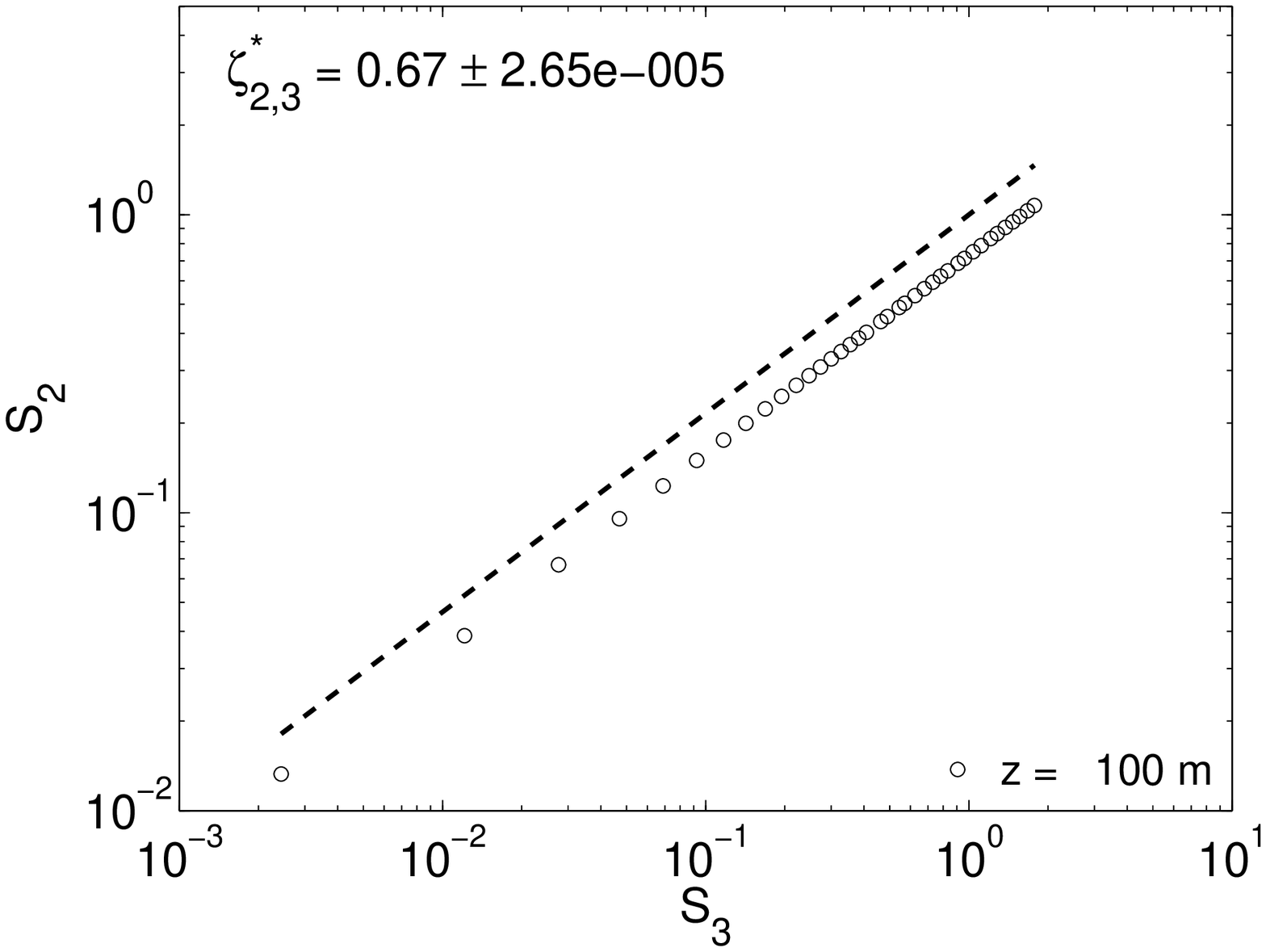}
\includegraphics[width=2.3in]{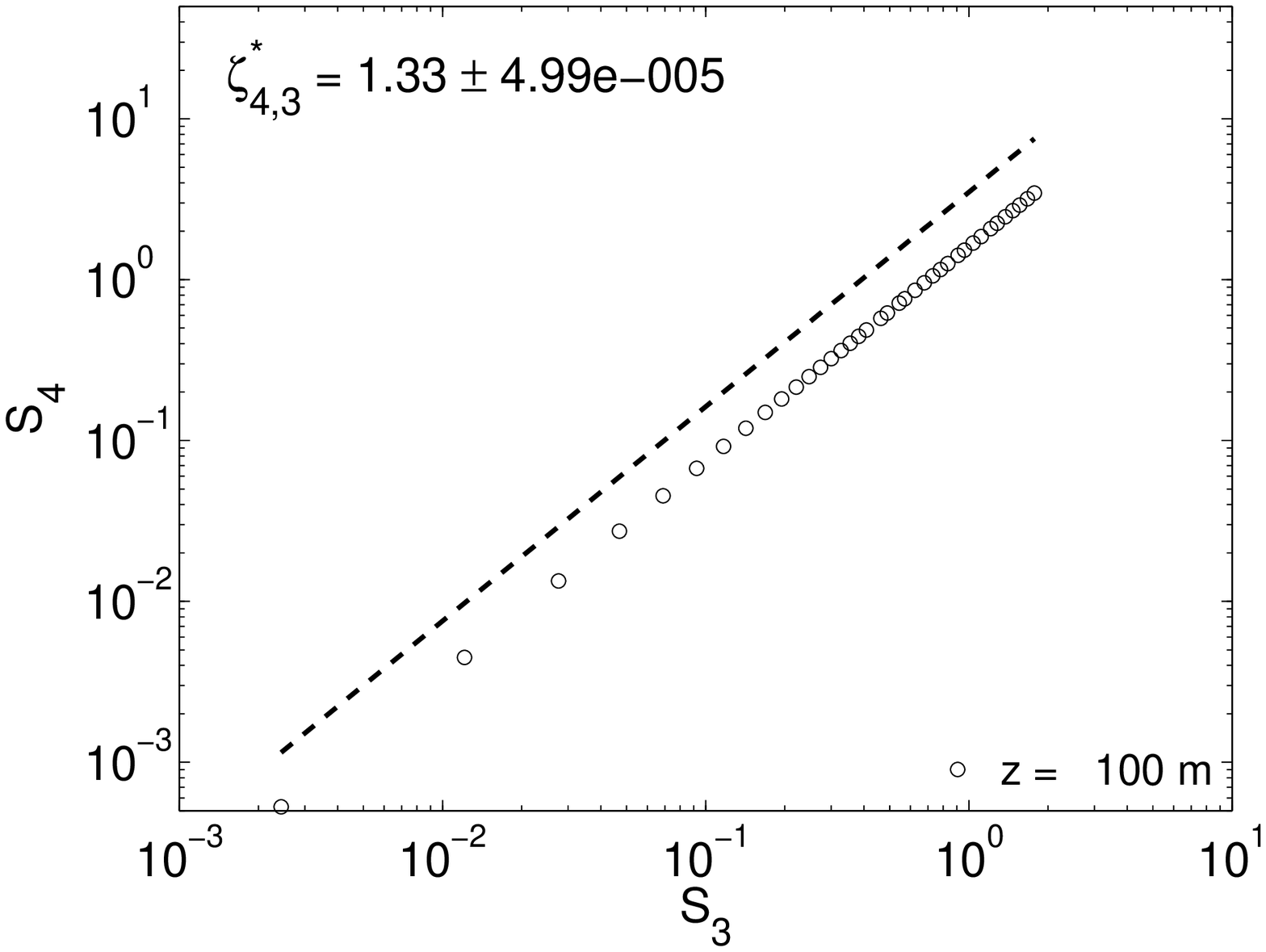}
\includegraphics[width=2.3in]{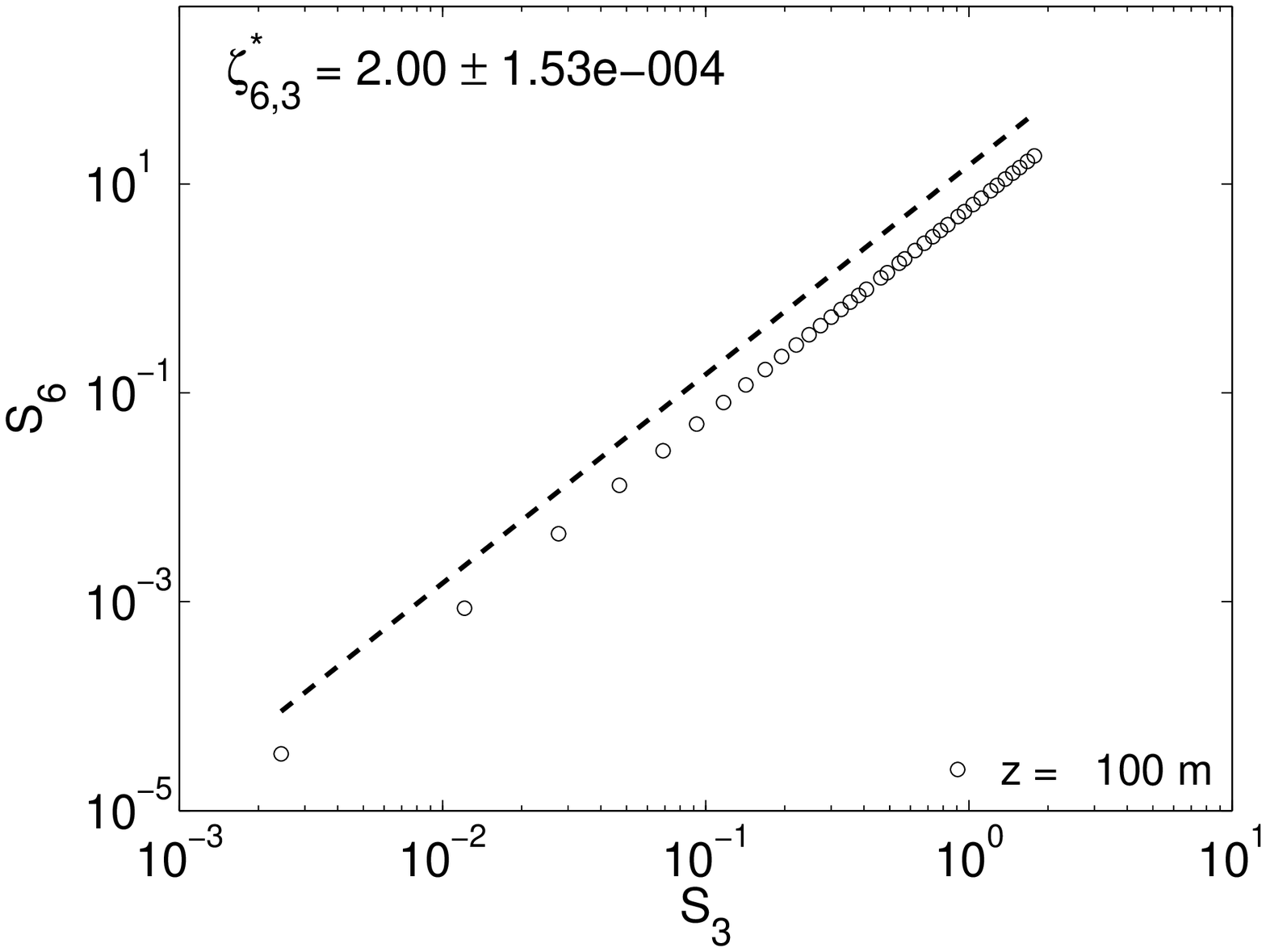}\\
\includegraphics[width=2.3in]{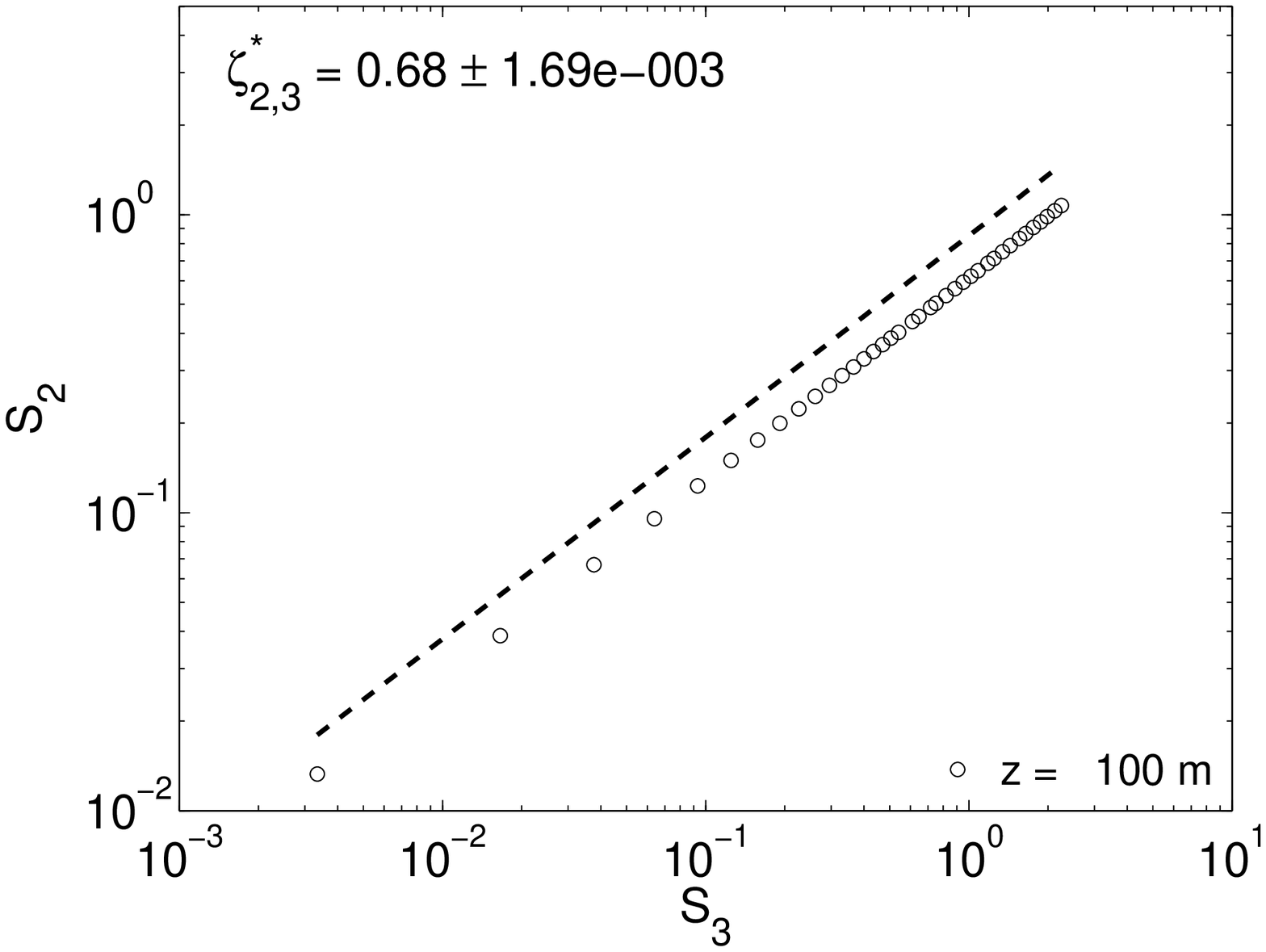}
\includegraphics[width=2.3in]{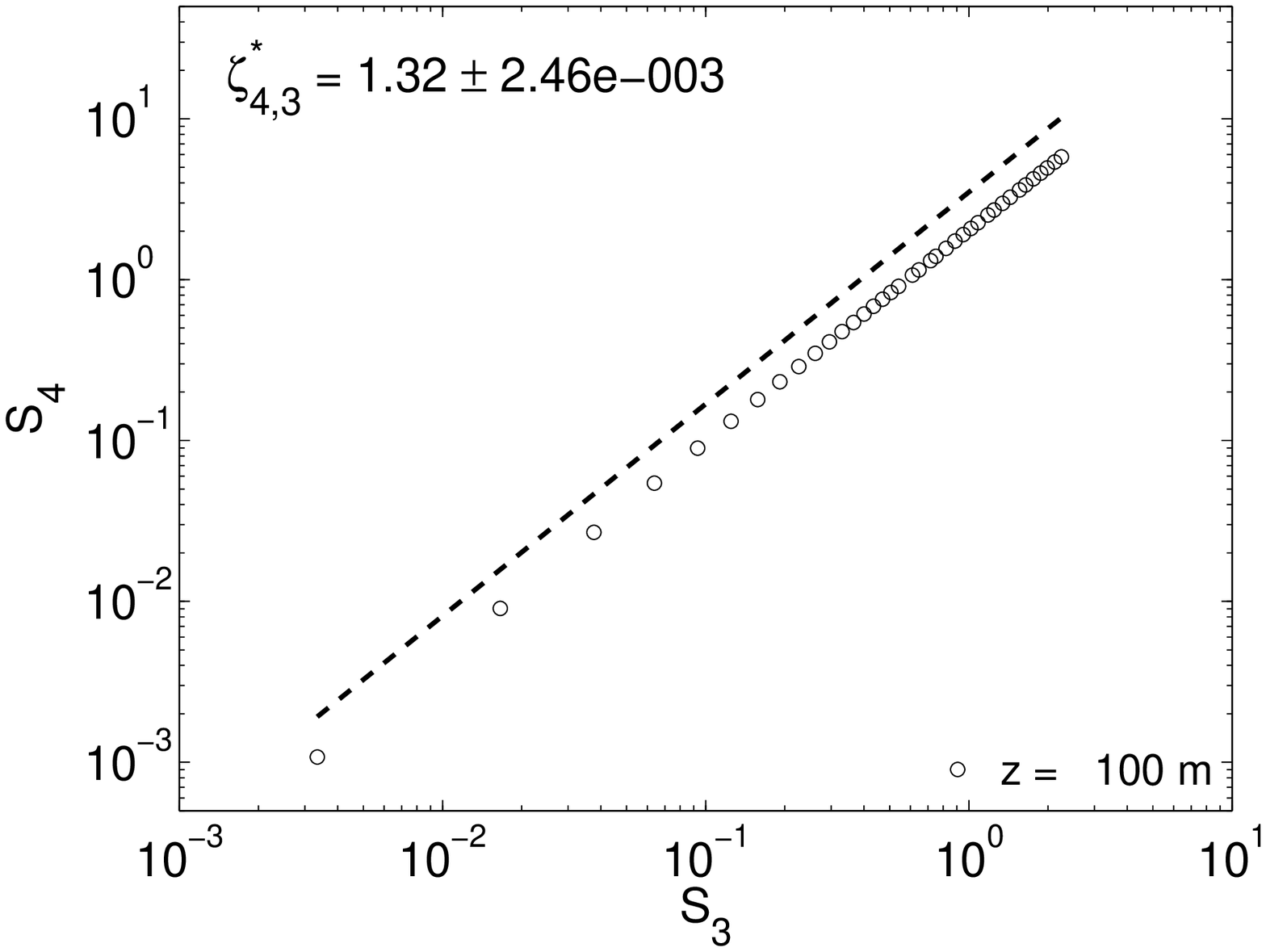}
\includegraphics[width=2.3in]{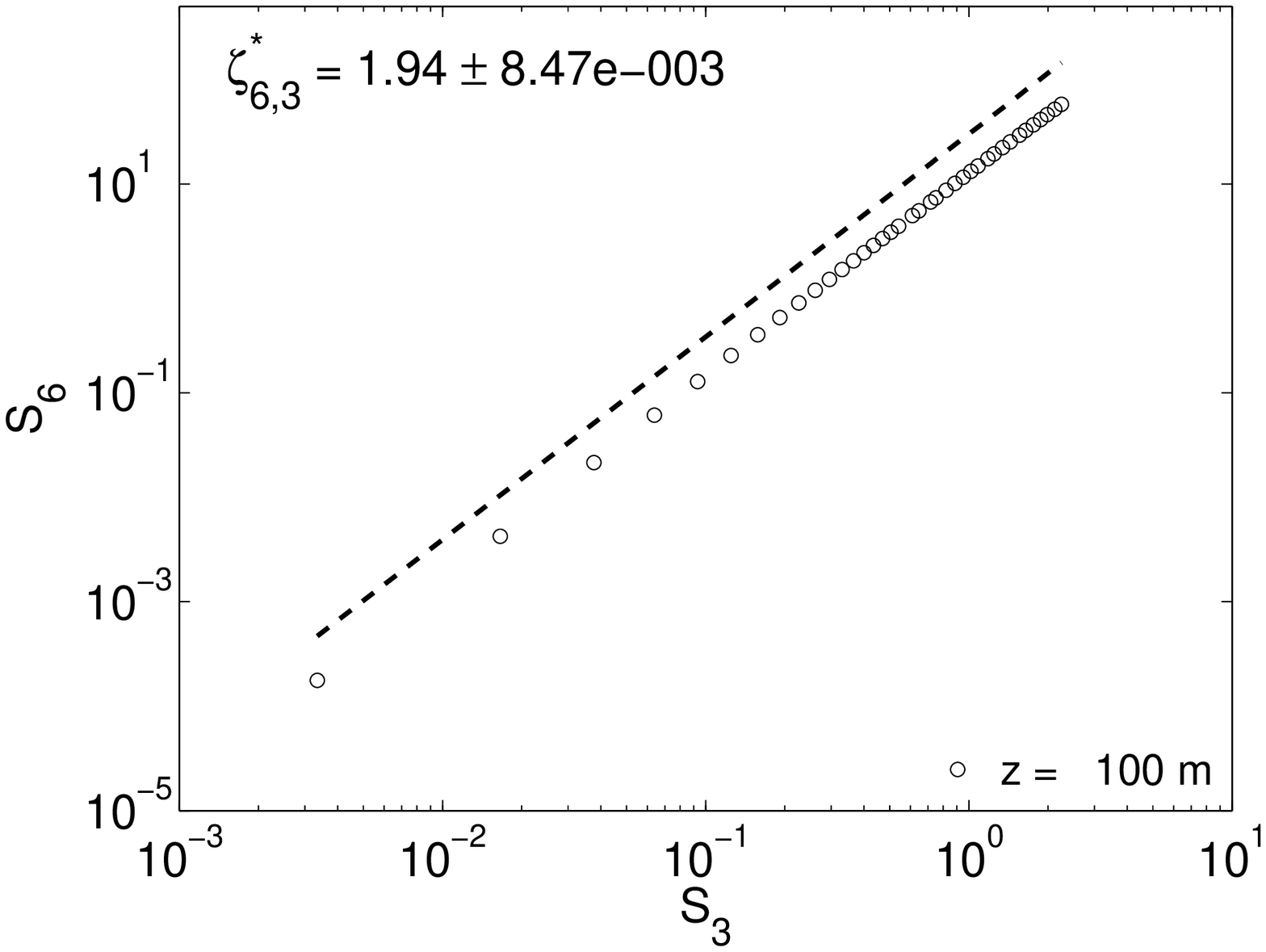}
\caption{The variation of the second-order (left panel), fourth-order (middle panel) and sixth-order (right panel) structure functions with respect to the third-order structure functions. The results from the FT surrogates of the observed and simulated wind series are shown in the first and third rows, respectively. Similar results from the IAAFT surrogates are presented in the second (observed) and fourth (simulated) rows. As in Fig.~\ref{f1}, the relative scaling exponents ($\zeta^*_{p,q}$) are reported on the top-left corner of each plots. The mean and standard deviation of $\zeta^*_{p,q}$ are estimated via bootstrapping. The dashed line in each plot represents the mean value of $\zeta^*_{p,q}$. For the first and second rows, the structure function values corresponding to $\Delta t = $ 1 min -- 6 h are used for $\zeta^*_{p,q}$ estimation. In contrast, for the third and fourth rows, the $\zeta^*_{p,q}$ values correspond to $\Delta t = $ 5 min -- 6 h.}
\label{f2}
\end{center}
\end{figure*}

By construction, the FT surrogates should portray ordinary scaling akin to monofractal series \cite{nikora2001extended}. Since the IAAFT surrogates are not truly linear (due to spurious phase correlations \cite{rath2012revisiting}), they might show marginal anomalous scaling behavior \cite{basu2007estimating,roux2009evidence}. Nevertheless, the ESS behavior of both types surrogates are expected to be quite different from their corresponding original time series. 
Various orders of structure functions are computed from each surrogate series and the ensemble averaged values are shown in Fig.~\ref{f2}. Like before, the scaling exponents are estimated via bootstrapping and reported in Table~\ref{T1} and Fig.~\ref{f2}. Clearly, the scaling exponents based on the FT surrogates are virtually indistinguishable from K-41 (i.e., $\zeta^*_{p,3} = p/3$). The ESS results based on the IAAFT surrogates are also very close to K-41. Thus, we can conclude that the anomalous scaling behaviors of the original wind series are not spurious; they are significantly different from corresponding surrogate (monofractal) series.        

In summary, we have provided empirical evidence that the ESS-based scaling holds for observed wind fields over complex terrain in the mesoscale regime. The scaling exponents are very close to the well-accepted inertial-range values. A state-of-the-art atmospheric model, with standard operational configuration, does not capture this scaling behavior. This shortcoming perhaps highlights the need for better turbulence closure parameterizations in new-generation atmospheric models. By virtue of the rigorous surrogate data analyses, it is prudent to say that all the reported results are statistically significant.

\acknowledgments
We are grateful to the National Renewable Energy Laboratory (NREL) for making the NWTC M2 and WIND Toolkit datasets publicly available for research. 

\providecommand{\noopsort}[1]{}\providecommand{\singleletter}[1]{#1}%

\end{document}